\documentclass[prx,aps,twocolumn,superscriptaddress]{revtex4-2}

\usepackage[pdftex]{graphicx}
\usepackage{amsbsy,amssymb,amsmath,bm,mathtools}
\usepackage{amsfonts}

\usepackage{xspace}
\usepackage{bm}
\usepackage{color}
 
\usepackage{url}
\usepackage[section]{placeins}
\usepackage[colorlinks=true,linkcolor=blue,citecolor=blue]{hyperref}

\begin{document}


\title{Coarsening of chiral domains in itinerant electron magnets: \\ A machine learning force field approach}

\author{Yunhao Fan}
\affiliation{Department of Physics, University of Virginia, Charlottesville, VA 22904, USA}

\author{Sheng Zhang}
\affiliation{Department of Physics, University of Virginia, Charlottesville, VA 22904, USA}

\author{Gia-Wei Chern}
\affiliation{Department of Physics, University of Virginia, Charlottesville, VA 22904, USA}

\date{\today}

\begin{abstract}
Frustrated itinerant magnets often exhibit complex noncollinear or noncoplanar magnetic orders which support topological electronic structures.  A canonical example is the anomalous quantum Hall state with a chiral spin order stabilized by electron-spin interactions on a triangular lattice.  While a long-range magnetic order cannot survive thermal fluctuations in two dimensions, the chiral order which results from the breaking of a discrete Ising symmetry persists even at finite temperatures. We present a scalable machine learning (ML)  framework to model the complex electron-mediated spin-spin interactions that stabilize the chiral magnetic domains in a triangular lattice. Large-scale dynamical simulations, enabled by the ML force-field models, are performed to investigate the coarsening of chiral domains after a thermal quench. While the chiral phase is described by a broken $Z_2$ Ising-type symmetry,  we find that the characteristic size of chiral domains increases linearly with time, in stark contrast to the expected Allen-Cahn domain growth law for a non-conserved Ising order parameter field. The linear growth of the chiral domains is attributed to the orientational anisotropy of domain boundaries. Our work also demonstrates the promising potential of ML models for large-scale spin dynamics of itinerant magnets.  
\end{abstract}

\maketitle

\section{Introduction}

Itinerant electron magnets offer a fruitful platform to study frustrated magnetism and complex spin textures~\cite{Hayami2021,Batista2016,Chern2015}. The coupling between local spins and itinerant electrons gives rise to long-ranged spin interactions and complex multi-spin couplings. The interplay between electron-mediated interactions and frustrated lattice geometry leads to complex magnetic structures that are characterized by noncollinar or noncoplanar spins in large extended unit cells. Perhaps the most prominent examples are magnetic crystals composed of topological spin textures such as vortices or skyrmions~\cite{Bogdanov1989,Roessler2006,Nagaosa2013,Seki2016,Tokura2020,Gobel2021}. In addition to their fundamental interest, the engineering and manipulation of such topological textures have important technological implications~\cite{Everschor2018,Fujishiro2020}.

The electronic structures such as the Fermi surface geometry play a crucial role in determining the effective spin interactions in itinerant magnets. In the limit of weak electron-spin coupling, the effective spin interaction is shown to be intimately related to the particle-hole susceptibility of the electron gas. For a spherical Fermi surface, which is a good approximation for most electron band structures at small filling fractions, this calculation leads to the famous Ruderman–Kittel–Kasuya–Yosida (RKKY) interaction~\cite{Ruderman1954,Kasuya1956,Yosida1957}. Itinerant magnets thus offer an avenue to realize complex magnetic textures via engineering of electronic structures. For example, a skyrmion crystal can also be viewed as a multiple-$\mathbf Q$ magnetic order driven by the nesting mechanism, with each ordering wave vector $\mathbf Q$ partially connecting two parallel segments of the Fermi surface~\cite{Martin2008,Chern2010,Hayami2014,Hayami2018,Wang2020,Hayami2021square}.  By manipulating the Fermi surface through, e.g. modification of hopping constants or filling fraction, one could control the structure and period of a skyrmion crystal.

The presence of a complex magnetic texture in itinerant magnets also endows electrons with a nontrivial Berry phase when they traverse around a closed loop of noncollinar or noncoplanar spins. This Berry phase effectively acts as a fictitious magnetic field which could reconstruct the electron band structure and enable nontrivial topological responses such as anomalous Hall effects~\cite{Onoda2004,Yi2009,Hamamoto2015,Gobel2017}. A canonical example of such topological magnetic states is the tetrahedral spin order on a triangular lattice~\cite{Momoi1997,Kurz2001,Martin2008}. This magnetic ordering has a quadrupled unit cell with local spins pointing toward corners of a regular tetrahedron, as shown in FIG.~\ref{fig:ml-scheme}.  In fact, the tetrahedral spin order with a uniform spin chirality can be viewed as the dense limit of skyrmion crystal with two skyrmions per magnetic unit cell. Moreover, the effective magnetic field generated by the tetrahedral order opens a gap at the Fermi surface and renders the magnet into a quantum Hall insulator~\cite{Martin2008,Chern2012}.

The noncoplanar spins in the tetrahedral order breaks the $Z_2$ chiral symmetry, which can be characterized by a nonzero scalar spin chirality $\chi = \langle \mathbf S_i \cdot \mathbf  S_j \times \mathbf S_k \rangle$ for three spins on an elementary triangular plaquette. As discussed above, the chiral ordered state exhibits a quantized Hall conductivity of $\pm e^2/h$, with the sign determined by that of the scalar chirality.
It is worth noting that the breaking of the discrete chiral symmetry in two dimensions persists at finite temperatures even though the static long-range magnetic order is destroyed by thermal fluctuations. The chiral phase transition is thus expected to belong to the 2D Ising universality class. However, the chiral ordering is shown by Monte Carlo simulations to be a discontinuous first-order transition~\cite{Kato2010,Barros2013}, indicating the unusual nature of this chiral symmetry breaking. 

Also of interest is the ordering dynamics of the chiral phase, which can be described by an Ising order parameter field in the coarse-grained approximation. The coarsening of Ising domains is one of the most studied subjects in the field of phase ordering dynamics~\cite{Bray1994,Onuki2002,Puri2009}. The evolution of the order-parameter fields when quenched into a symmetry-breaking phase is often highly nonlinear and is characterized by the emergence of complex spatial patterns. Depending on whether the Ising order is conserved, the growth of Ising domains is governed by well-known super-universality classes. For non-conserved Ising order, as in the case of chiral order associated with the tetrahedral spin structure, its coarsening is expected to follow the Allen-Cahn power law~\cite{Allen1972} $L(t) \sim \sqrt{t}$, where $L(t)$ is the characteristic linear size of Ising domains. 

Yet, to the best of our knowledge, the coarsening of chiral domains in itinerant magnets has yet been carefully examined. This is partly due to the difficulty of large-scale dynamical simulations of itinerant magnets. In conventional Landau-Lifshitz-Gilbert (LLG) simulations of spin dynamics, the local magnetic fields acting on spins can be efficiently computed from an effective spin Hamiltonian with, e.g. short-ranged Heisenberg exchange interactions. On the other hand, since the driving force of spins in itinerant magnets comes from electrons, the calculation of local exchange field requires solving an electronic structure problem which has to be repeated at every time-step of dynamical simulations. For itinerant spin models with free electrons, the electron bands can be solved by exact diagonalization. However, even for such relatively simpler systems, repeated diagonalization would be prohibitive for large-scale dynamical simulations due to the $\mathcal{O}(N^3)$ time complexity of the method, where $N$ is the number of spins.  

In this paper, we present a linear-scaling machine learning (ML) framework to investigate the coarsening of chiral domains in itinerant frustrated magnets. The scalability of our ML approach relies fundamentally on the locality principle, which in our case means the magnetic field acting on a local spin depends predominantly on its immediate neighborhood. A deep-learning neural network (NN) model is developed to encode the complicated dependence of the exchange field on the neighborhood spin configurations. Importantly, since the NN takes a finite neighborhood as the input, the resultant ML force-field model is both transferrable and scalable, which means that the same ML model, successfully trained from small-scale exact solutions, can be applied to much larger systems without rebuilding or retraining.

We apply the ML approach to the spin dynamics of the triangular Kondo-lattice model, a canonical system of frustrated itinerant magnets. The model exhibits the tetrahedral magnetic order accompanied by a quantized Hall electronic state at zero temperature. Large-scale LLG simulations enabled by the ML force-field model uncover an intriguing coarsening behavior where the characteristic domain size increases nearly linearly with time at late stage of the phase ordering. This anomalous coarsening of chiral domains which is faster than the expected Allen-Cahn behavior is attributed to the anisotropy, both in shape and orientation, of the interfaces that separates two chiral domains of opposite sign.

The rest of the paper is organized as follows. In Sec.~\ref{sec:ML}, we present a scalable ML force field model for itinerant electron magnets. The ML model is benchmarked against the kernel polynomial method for the force prediction and time-dependent spin-spin correlation functions.  We then briefly discuss the structure and properties of the chiral tetrahedral spin order in the triangular Kondo-lattice model in Sec.~\ref{sec:tetra-order}.  The ML method is then applied to enable large-scale thermal quench simulations of the Kondo-lattice model. Detailed characterizations of the phase ordering dynamics and the growth law of chiral domains are presented in Sec.~\ref{sec:results}. Finally, we conclude in Sec.~\ref{sec:discussion} with a summary and outlook.

\section{Machine learning force fields for itinerant electron magnets}

\label{sec:ML}

\subsection{Landau-Lifshitz-Gilbert dynamics for itinerant electron magnets}

We consider a one-band Kondo-lattice model (KLM), also known as the s-d model, as a representative system for itinerant magnets. Its Hamiltonian reads
\begin{eqnarray}
	\label{eq:H1}
	\hat{\mathcal{H}} =- \sum\limits_{ ij , \alpha} t_{ij}  \left(\hat{c}_{i \alpha}^{\dagger}\hat{c}^{\,}_{j \alpha} + \mbox{h.c.} \right)
	+J \sum_{i, \alpha\beta} \mathbf {S}_i \cdot \hat{c}^\dagger_{i\alpha} \bm\sigma^{\,}_{\alpha\beta} \hat{c}^{\,}_{i \beta},  \quad
\end{eqnarray}
where $\hat{c}_{i \alpha}^{\dagger}(\hat{c}^{\,}_{i \alpha})$ is the creation (annihilation) operator of an electron with spin $\alpha = \uparrow, \downarrow$ at site-$i$. The first  $t_{ij}$ term denotes nearest-neighbor pairs on the triangular lattice, while $J$ in the second term represents the Kondo coupling between local magnetic moment $\mathbf S_i$ and electron spin, and $\bm \sigma_{\alpha\beta}$ denotes the vector of Pauli matrices.
Since we are interested in the magnetization dynamics, we approximate local moments as 3-vectors of a constant length. For convenience, we use a unit system such that the local moments are dimensionless vectors with a unit length~$|\mathbf S_i| = 1$.

The magnetization dynamics for itinerant magnets is governed by the Landau-Lifshitz-Gilbert (LLG) equation
\begin{eqnarray}
	\label{eq:LLG}
	\frac{d\mathbf S_i}{dt} = \gamma \mathbf S_i \times ( \mathbf H_i + \bm\eta_i )  
	- \alpha \mathbf S_i \times (\mathbf S_i \times \mathbf H_i).
\end{eqnarray}
Here  $\gamma$ is the gyromagnetic ratio, $\alpha$ is the damping constant, $\mathbf H_i$ is the local effective field, and $\bm\eta_i$ represents a stochastic magnetic field due to thermal fluctuations. The thermal fields at different sites are independent of each other and their time-dependence is modeled by a white noise with a zero mean. The driving force of the LLG equation in the adiabatic approximation is assumed to be conservative and is given by the derivative of an effective energy of the system:
\begin{eqnarray}
	\label{eq:local-H}
	\mathbf H_i = -\frac{\partial E}{\partial \mathbf S_i}.
\end{eqnarray}
This total energy $E = E(\{\mathbf S_i \})$ as a function of spins can thus be viewed as an effective classical spin Hamiltonian obtained by integrating out the electron degrees of freedom. In the weak coupling limit $J \ll t_{ij}$, a second-order perturbation calculation gives an effective spin Hamiltonian $E = \sum_{ij} \mathcal{J}(|\mathbf r_i - \mathbf r_j|) \mathbf S_i \cdot \mathbf S_j$, with a long-range interaction $\mathcal{J}(r)$ similar to the well-known RKKY interaction~\cite{Ruderman1954,Kasuya1956,Yosida1957}.
For intermediate and large electron-spin coupling, however, higher-order terms in the expansion cannot be neglected. In fact, both biquadratic and six-order terms are shown to play an important role in the stabilization of the chiral tetrahedral order~\cite{Momoi1997,Akagi2012}. Systematic inclusion of higher order terms, however, is rather tedious, if not impossible. 

The effective spin Hamiltonian can be explicitly computed within the adiabatic approximation, which assumes a much faster electronic relaxation compared with the spin dynamics. This is similar to the Born-Oppenheimer approximation in quantum molecular dynamics (MD)~\cite{Marx09}.  Assuming a quasi-equilibrium electron subsystem, the effective energy is given by the expectation value of the Hamiltonian
\begin{eqnarray}
	\label{eq:E_quantum}
	E = \bigl\langle \hat{\mathcal{H}}(\{\mathbf S_i\}) \bigr\rangle 
	= {\rm Tr}\bigl(\hat{\rho}_e \hat{\mathcal{H}}  \bigr),
\end{eqnarray}
where $\hat{\rho}_e = \exp(-\beta \hat{\mathcal{H}})/\mathcal{Z}$ is the equilibrium electron density matrix corresponding to an instantaneous spin configuration, and $\mathcal{Z} = {\rm Tr}{\hat{\rho}_e}$ is the partition function. For a given spin configuration $\{\mathbf S_i \}$, the effective energy can be obtained from the exact diagonalization (ED) of the quadratic electron Hamiltonian in Eq.~(\ref{eq:H1}). However, since the time complexity of ED scales as $\mathcal{O}(N^3)$, where $N$ with the number of spins, large-scale LLG simulation based on the ED would be prohibitively time-consuming. 

The Kondo-lattice Hamiltonian~(\ref{eq:H1}) can also be solved using more efficient methods. Notably,  the kernel polynomial methods (KPM) offer a linear-scaling approach to computing spectral properties of large-scale quadratic Hamiltonians~\cite{Weisse06,Barros13,Wang18}. For example, by expanding the density of states (DOS) of the Hamiltonian as a polynomial series, $\varrho(\epsilon) = \sum_m c_m T_m(\epsilon)$, where $T_m(x)$ are Chebyshev polynomials, the expansion coefficients $c_m$ can be efficiently and recursively computed based on sparse-matrix vector multiplication. The total energy can then be obtained as $E = N \int d\epsilon \, \varrho(\epsilon)$. For LLG dynamics, the central task is the computation of local effective field: $\mathbf H_i = -\langle  \partial \hat{\mathcal{H}} / \partial \mathbf S_i \rangle = -J \bm \sigma^{\,}_{\alpha\beta} \rho^{\,}_{i\beta, i\alpha}$, where the single-particle density matrix $\rho^{\,}_{j\beta, i\alpha} \equiv \langle \hat{c}^\dagger_{i, \alpha} \hat{c}^{\,}_{j,\beta} \rangle$ can also be efficiently obtained using a similar Chebyshev series expansion. In this work, the KPM is used to generate the datasets and to serve as the benchmark for the ML models.

\subsection{Behler-Parrinello machine learning models}

\label{sec:BP}

Machine learning (ML) methods offer a general linear-scaling approach to the computation of effective local fields $\mathbf H_i$. Fundamentally, as pointed out by W. Kohn, linear-scaling electronic structure methods are possible mainly because of the locality nature or ``nearsightedness'' principle of many-electron systems~\cite{kohn96,prodan05}.   The nearsightedness of electron systems here does not rely on the existence of well localized Wannier-type wave functions, which only exist in large-gap insulators. Instead, the locality is generally a consequence of wave-mechanical destructive interference of many-particle systems. An important implication of the locality principle is that extensive physical quantities can be computed via a divide-and-conquer approach. 

Indeed, the locality principle is the cornerstone of recent ML-based force-field methods that enable large-scale {\em ab initio} MD simulations with the desired quantum accuracy~\cite{behler07,bartok10,li15,shapeev16,behler16,botu17,smith17,chmiela17,zhang18,chmiela18,deringer19,mcgibbon17,suwa19,sauceda20}. In these approaches, the atomic forces, which play a central role in MD simulations, are assumed to depend on the local chemical environment. A ML model of a fixed size is developed to encode the highly complicated dependence of atomic force on the local neighborhood. A practical implementation of such ML force-field models was demonstrated in the pioneering works of Behler and Parrinello~\cite{behler07} and Bart\'ok {\em et~al.}~\cite{bartok10}. Instead of directly predicting the atomic forces, the ML model is trained to produce so-called atomic energy from the local chemical environment. The atomic forces are obtained indirectly from the total energy, which is the sum of all atomic energies. As will be discussed in more detail below, one important advantage of this approach is that symmetry properties of the original quantum systems can be readily incorporated into the effective ML models.  Similar ML frameworks have recently been developed to enable large-scale dynamical simulations in several condensed-matter lattice systems~\cite{zhang20,zhang21,zhang22,zhang22b,zhang23,cheng23,cheng23b}.

\begin{figure*}[t]
\centering
\includegraphics[width=1.99\columnwidth]{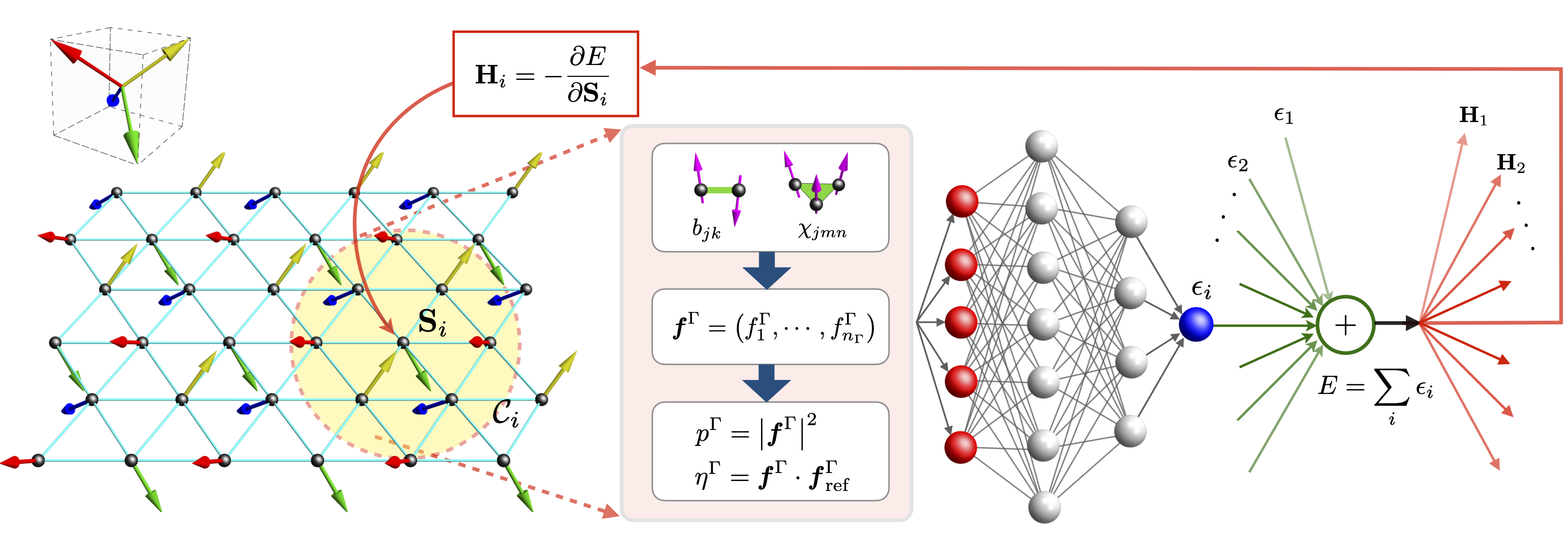}
\caption{Schematic diagram of ML force-field model for itinerant electron magnets. A descriptor transforms the neighborhood spin configuration $\mathcal{C}_i$ to effective coordinates $\{ p^\Gamma_r, \eta^\Gamma_r \}$ which are then fed into a neural network (NN). The output node of the NN corresponds to the local energy $\epsilon_i = \varepsilon(\mathcal{C}_i)$ associated with spin $\mathbf S_i$. The corresponding total potential energy $ E$ is obtained from the summation of these local energies. Automatic differentiation is employed to compute the derivatives $\partial E / \partial \mathbf S_i$ from which the local exchange fields $\mathbf H_i$ are obtained. }
    \label{fig:ml-scheme}
\end{figure*}

Here we present a Behler-Parrinello (BP) ML framework to enable large-scale LLG dynamics simulations for itinerant electron magnets. A schematic diagram of the ML force-field model is shown in FIG.~\ref{fig:ml-scheme}. First, the effective energy $E$ defined in Eq.~(\ref{eq:E_quantum}), i.e. obtained by integrating out the electrons, is decomposed into local energies $\epsilon_i$, each associated with a lattice site:
\begin{eqnarray}
	\label{eq:E_ML}
	E = \sum_i \epsilon_i = \sum_i \varepsilon(\mathcal{C}_i).
\end{eqnarray}
Next, as expressed in the second equality above, we invoke the locality principle and assume that the site energy $\epsilon_i$ depends only on a local magnetic environment, denoted as $\mathcal{C}_i$, through a universal function $\varepsilon(\cdot)$ which is determined by the underlying electron Hamiltonian. In practical implementations, the neighborhood is defined as the collection of spins within a cutoff radius~$r_c$ around the $i$-th site: $\mathcal{C}_i = \{ \mathbf S_j \,  \bigl| \, |\mathbf r_j - \mathbf r_i| < r_c \}$. The complex functional dependence of the local energy on the magnetic environment is to be approximated by a learning model. 

In ML-based MD methods, several ML schemes have been proposed to represent the atomic energy as a function of local chemical or atomic environment. Notable among them are the Gaussian approximation potential~\cite{bartok10} and the neural-network potential~\cite{behler07,zhang18}. Here we use feedforward neural networks (NN) as the learning model which, according to universal approximation theorem~\cite{Cybenko89,Hornik89}, offers the capability of accurately representing complex functions to the desired accuracy.

Given a successfully trained ML model for the local energy, the total energy $E$ is obtained by summing over all local energies obtained by applying the same ML model to every single spin in the lattice. The fact that the same ML model is used for all lattice sites simply reflects the translational symmetry of the original Hamiltonian. The overall computation is formally equivalent to stacking $N$ identical ML models together to form a super neural network which takes the whole spin configuration $\{\mathbf S_i\}$ as input and produces the total energy $E$ at the output node; see FIG.~\ref{fig:ml-scheme}. The effective local field $\mathbf H_i$, given by the derivative of the total energy shown in Eq.~(\ref{eq:local-H}), can be efficiently computed using the automatic differentiation techniques~\cite{Paszke17,Baydin18}. Practically, however, thanks to the locality of the effective field, the calculation of $\mathbf H_i$ requires only a partial energy that includes local energies $\epsilon_j$ in a finite neighborhood around site-$i$, instead of contribution from the whole lattice, thus ensuring that the approach is linear-scaling.

\subsection{Magnetic descriptor: Representation of local magnetic environment}

\label{sec:m-descriptor}

A crucial part of the ML force-field model is the proper representation, also known as a descriptor, of spin configurations $\mathcal{C}_i$ in a local neighborhood. This is particularly important for the BP type schemes since the local energy~$\epsilon_i$ at the output, as a scalar, is invariant under the symmetry operations of the system. Consequently, the descriptor of the local neighborhood not only should be able to differentiate distinct spin-configurations, but also remain invariant under symmetry transformations of the original electron model. Indeed, descriptors also play a central role in almost all ML-based force field models for quantum MD simulations. In the case of molecular systems, the proper representation must respect basic symmetries including rotation, reflection, and permutation of same-species atoms.  

Over the past decade, several atomic descriptors have been proposed together with the learning models based on them~\cite{behler07,bartok10,li15,behler11,ghiringhelli15,bartok13,drautz19,himanen20,huo22}. For example, the atom-centered symmetry functions (ACSF), which is one of the most popular atomic descriptors, is based on the two-body (relative distances) and three-body (relative angles) invariants of an atomic configuration~\cite{behler07,behler16}. The group-theoretical method, on the other hand, offers a more controlled approach to the construction of atomic representation based on the bispectrum coefficients~\cite{bartok10,bartok13}. 

The group-theoretical method has also been employed to develop a general theory of descriptors for electronic lattice models in condensed-matter systems~\cite{zhang22,Ma19,Liu22,Tian23}. Compared with MD systems, the SO(3) rotational symmetry of free-space is reduced to discrete point-group symmetries in lattice models. On the other hand, the dynamical degrees of freedom in lattice models, such as local magnetic moments or order-parameters, are characterized by additional internal symmetry group. A proper descriptor for lattice systems thus needs to be invariant with respect to both the internal symmetry group and the lattice point group. In particular, magnetic descriptors based on group-theoretical bispectrum methods have recently been constructed for itinerant electron magnets~\cite{zhang21,zhang23}. Here we outline the theory and procedure for constructing the feature variables from the magnetic environment.

The Kondo-lattice Hamiltonian in Eq.~(\ref{eq:H1}) is invariant under two independent symmetry groups: the SO(3)/SU(2) rotation of spins and the point group of the site-symmetry of the underlying lattice. The local energy function $\varepsilon(\mathcal{C}_i)$ should preserve these two sets of symmetry operations.  First, the spin rotation symmetry can be manifestly maintained if the energy function only depends on the bond variables $b_{jk}$ and scalar chirality $\chi_{jmn}$ of spins within the neighborhood 
\begin{eqnarray}
	\label{eq:bond-chirality}
	b_{jk} = \mathbf S_j \cdot \mathbf S_k, \qquad \chi_{jmn} = \mathbf S_j \cdot \mathbf S_m \times \mathbf S_n.
\end{eqnarray}
These two building blocks also correspond to two-spin and three-spin correlations. 
Next, we consider discrete site symmetries of the lattice, which is described by point group $D_6$ in our case. The collection of bond and chirality variables $\{b_{jk}, \chi_{jmn} \}$~in the neighborhood of the $i$-th site form a basis of a high-dimensional representation of the $D_6$ group. They can then be decomposed into fundamental irreducible representations (IR’s) of the point group. For example, consider the six bond variables $b_\mu = \mathbf S_i \cdot \mathbf S_\mu$, formed from a center spin $\mathbf S_i$ and its six neighbors $\mathbf S_\mu$ ($\mu = A, B, \cdots, F$). These six bond variables can be decomposed into one $A_1$, one $B_1$, and two doublet $E$ IRs. The coefficients or basis of these IRs are orthogonal transformations of the bond variables, e.g. the coefficient of the $A_1$ IR is given by the symmetric sum: $f^{A_1} = (b_A + b_B + \cdots + b_F)$. More details of the IR calculations are discussed in Appendix~\ref{sec:descriptor}. 

For convenience, we arrange the coefficients of a given IR into a vector $\bm f^{(\Gamma, r)} = (f^{(\Gamma, r)}_1, f^{(\Gamma, r)}_2, \cdots, f^{(\Gamma, r)}_{n_\Gamma} )$, where $\Gamma$ denotes the type of IR, $n_\Gamma$ its dimension, and $r$ enumerates the multiplicity of the IR.  One immediate class of invariants is their amplitudes 
\begin{eqnarray}
	\label{eq:power-spectrum}
	p^\Gamma_r=|\boldsymbol{f}^\Gamma_r|^2, 
\end{eqnarray}
which is called the power spectrum of the representation. However, descriptors based solely on the power spectrum ignore the relative phases between different IRs.  A more general set of invariants which include relative phase information is the so-called bispectrum coefficients~\cite{kondor07,bartok13}. A bispectrum coefficient is a product of three IR coefficients and the Clebsch-Gordon coefficients which account for the different transformation properties of the three IRs. For most point groups, the dimensions $D_\Gamma$ of individual IRs are small, which means there is a large multiplicity (indexed by $r$) for each IR. This in turn results in a large number of possible bispectrum coefficients, often with considerable redundancy. 

In order to keep the number of feature variables manageable, we introduce the concept of reference IR coefficients $\bm f^\Gamma_\text{ref}$, one for each IR type of the point group~\cite{zhang21}. These reference IR coefficients are derived using the same decomposition formulas, but based on variables $\overline{b}_\mu$ and $\overline{\chi}_\mu$ obtained by averaging large blocks of bond and chirality variables in the local neighborhood in order to reduce sensitivity to small variations. Importantly, the reference IR allows one to introduce a ``phase" variable for each IR in the decomposition 
\begin{eqnarray}
	\label{eq:IR-phase}
	\eta^\Gamma_r \equiv \bm f^\Gamma_r \cdot \bm f^\Gamma_{\rm ref} / |\bm f^\Gamma_r |\, |\bm f^\Gamma_{\rm ref}|. 
\end{eqnarray}	
The relative phase between IRs of the same type can then be inferred from their respective phases relative to the reference. The final feature variables consist of the amplitude and phase variables $\{p^\Gamma_r, \eta^\Gamma_r\}$ along with a finite number of bispectrum coefficients of the reference IRs. Importantly, the site energy $\epsilon_i$ produced by the output of the NN depends on the local environment only through these invariant feature variables, hence preserving the symmetry of the original KLM.

\subsection{Benchmark of ML force-field models}

Here we present benchmark results of the ML model applied to the chiral phase of the triangular KLM. The chiral phase with an underlying tetrahedral spin order is stabilized by an electron-spin coupling $J = 3$ and a chemical potential $\mu = -3.2$, corresponding to a filling fraction $n \approx 1/4$; here both energies are measured in units of the nearest-neighbor electron transfer $t_{\rm nn}$. The ML force predictions are compared against the results obtained from KPM~\cite{Weisse06,Barros13,Wang18}. To further improve the efficiency of the LLG simulation with KPM, an automatic differentiation with reverse accumulation method~\cite{Griewan89} is used to facilitate the force calculation and a probing method~\cite{Tang12} is used to speed up the multiplications of sparse matrix and vectors~\cite{Barros13,Wang18}. The number of Chebyshev polynomials used in the simulations is in the range of $M = 1000$ to 2000, and the number of correlated random vectors used in the probing method is $R = 64$ to 144.

Next, we discuss the implementation details of the ML force model. First, a cutoff radius $r_c = 6 a$, where $a$ is the lattice constant of the triangular lattice, is used in defining the neighborhood $\mathcal{C}_i$. Within a neighborhood, we only include off-center pairs $(mn)$ with a separation $r_{mn} < 2a$ for the bond $b_{mn}$ and chirality variables $\chi_{imn}$. A total of 1806 bond and chirality variables are included as the building blocks to produce the same number of feature variables $ \{ p^\Gamma_r, \eta^\Gamma_r \} $, which are the input to the neural network. 
The neural network model, implemented using PyTorch~\cite{Paszke2019}, comprises eight hidden layers. The input layer of the model was determined by the number of feature variables, which in this case is 1806. The only output node produces the local energy $\epsilon_i$.  Details of the NN structure, the loss function, the training dataset, and optimizer are discussed in Appendix~\ref{sec:nn-model}. 

\begin{figure}[t]
\centering
\includegraphics[width=0.99\columnwidth]{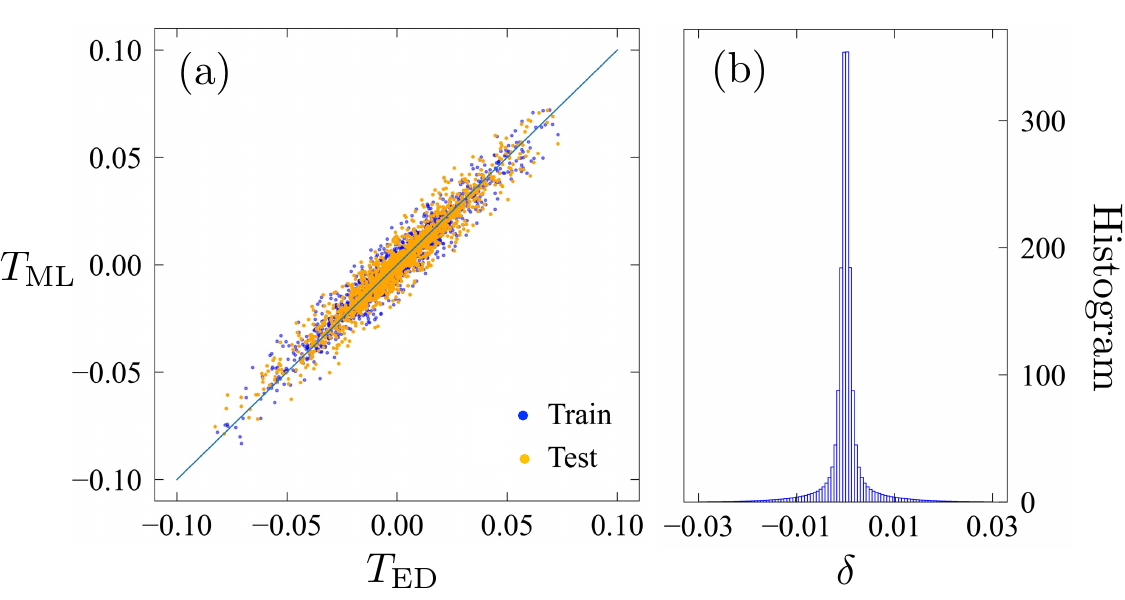}
\caption{Benchmark of ML for the adiabatic dynamics of the KLM. Panel (a) compares components of the ML-predicted torque $T_{\rm ML}$ and the ground truth $T_{\rm ED}$ obtained from exact diagonalization (ED). The torque is defined as $\mathbf T_i = \mathbf S_i \times \mathbf H_i$, where $\mathbf H_i$ is the local effective field for the $i$-th spin. Panel (b) shows the histograms of the prediction error~$\delta =T_{\rm ED} - T_{\rm ML}.$ }
\label{fig:benchmark-force}
\end{figure}

FIG.~\ref{fig:benchmark-force}(a) shows components of the torques $\mathbf T_i = \mathbf S_i \times \mathbf H_i$ predicted by the ML model versus the ground truth for both the training and testing data. A MSE of $1.4 \times 10^{-5}$ is obtained for the training dataset. A similar MSE $1.64\times 10^{-5}$ is computed from the testing data; the balanced outcome between training and testing datasets indicates that the ML model does not suffer from overfitting.  The normalized distribution of the prediction error obtained from the validation dataset, shown in the inset of FIG.~\ref{fig:benchmark-force}(b), is characterized by a rather small standard deviation of $\sigma = 0.004$. Overall, the ML-predicted forces agree very well with the ground truth.

The ML model is then integrated with the LLG method to simulate thermal quench of the KLM for dynamical benchmarks. Essentially,  the above trained ML model is employed to compute the effective local field $\mathbf H_i$ at every time-step of the dynamical simulation. The results are then compared against LLG simulations based on KPM for the computation of local fields. The simulation time here is measured in unit of the timescale associated with the electron hopping $\tau_0 = (\gamma t_{\rm nn})^{-1}$, where $\gamma$ is the gyromagnetic ratio and $t_{\rm nn}$ is the nearest-neighbor hopping coefficient. In the following, a time-step $\Delta t = 0.025 \tau_0$ and a dimensionless dissipation coefficient $\alpha = 0.075$ are used in all simulations.

\begin{figure}[t]
\centering
\includegraphics[width=0.99\columnwidth]{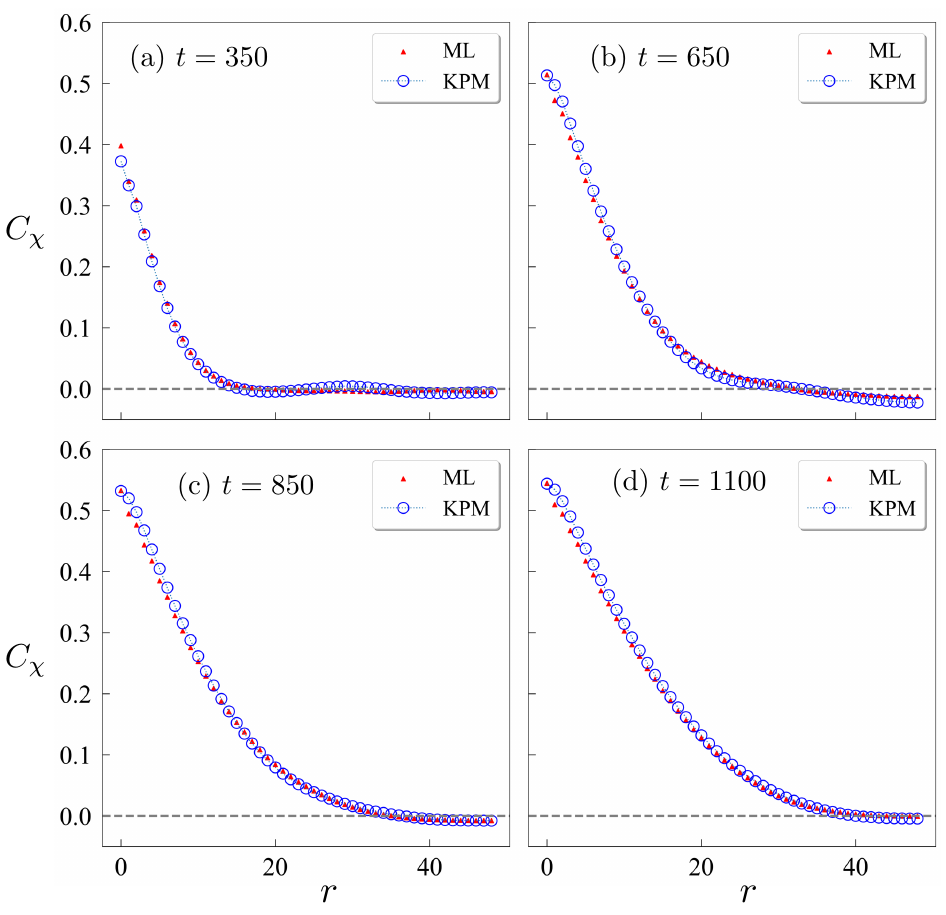}
\caption{Chirality correlation functions $C_{\chi}(\mathbf r)$ at various times after a thermal quench at $t = 0$. These results are obtained from LLG simulations on a $96\times 96$ lattice. The red solid and blue open circles are data points with local fields computed using the ML force-field model and the KPM, respectively.}
    \label{fig:correlation}
\end{figure}

For thermal quench simulations, the spins are coupled to a thermal bath of temperature $T$ through the Langevin thermostat. The system is initially prepared in a state of random spins, corresponding to an equilibrium at infinite temperature. The thermal bath is suddenly quenched to a low-temperature $T = 10^{-4}\,t_{\rm nn}$  at time $t = 0$. As this temperature is well below the critical temperature of the chiral phase transition, the subsequent relaxation of the system is dominated by the development of long-range chiral order and the quasi-long-range tetrahedral spin order. 
To quantify the development of the chiral order, we consider the following chirality correlation function 
\begin{eqnarray}
	C_{\chi}(\mathbf r, t) = \langle \delta\chi_{\triangle}(\mathbf r_0, t) \delta\chi_{\triangle}(\mathbf r_0 + \mathbf r, t) \rangle, 
\end{eqnarray}
where $\delta\chi_{\triangle}(\mathbf r) = \chi_{\triangle}(\mathbf r) - \langle \chi_{\triangle} \rangle$, and $\chi_{\triangle}(\mathbf r) = \mathbf S_1 \cdot \mathbf S_2 \times \mathbf S_3$ denotes the spin chirality of a triangle centered at $\mathbf r$ and $\mathbf S_{1, 2, 3}$ are the three spins on the triangle. The brackets $\langle \cdots \rangle$ indicate the average over the reference lattice point $\mathbf r_0$ and independent runs from different initial conditions. FIG.~\ref{fig:correlation} shows the $C_{\chi}(\mathbf r, t)$ functions, obtained from both KPM and ML-LLG simulations, at various times after the thermal quench.  The correlation functions obtained from the two approaches agree well with each other. This dynamical benchmark indicates that the ML model not only accurately predicts the local fields, but also captures the dynamical evolution of the Kondo-lattice system. 

As shown in Fig.~\ref{fig:correlation}, the chirality correlation after the quench decays with increasing separation $|\mathbf r|$, indicating a finite correlation length $\xi$. As will be discussed in more detail in Sec.~\ref{sec:results}, domains of well-defined local tetrahedral spin order quickly develop after a thermal quench. As a result, the finite $\xi$ results from the relative small sizes of such chiral domains after the quench. The gradual enhancement of the chirality correlation length with increasing time is due to the growth of the chiral domains.

\section{Chiral tetrahedral spin order on triangular lattice}

\label{sec:tetra-order}

Here we review the properties of the chiral tetrahedral spin order in the triangular KLM. This itinerant spin system shows an array of interesting magnetic phases depending on the electron-spin coupling $J$ and the electron filling fraction $n$~\cite{Akagi10,Azhar17}. Of particular interest to our work is the chiral tetrahedral spin order which is stabilized by a small $J$ at filling fractions $n \sim 1/4$ and $3/4$. This magnetic order has a quadrupled unit cell with local spins pointing toward the corners of a regular tetrahedron, as shown in the left panel in FIG.~\ref{fig:ml-scheme}. In general, a magnetic order with a quadrupled unit cell is related to a triple-$\mathbf Q$ magnetic structure that is characterized by three vector order parameters $\bm\Delta_\eta$:
\begin{eqnarray}
	\label{eq:triple-Q}
	\mathbf S_i = \bm \Delta_1 e^{i \mathbf Q_1 \cdot \mathbf r_i} + \bm \Delta_2 e^{i \mathbf Q_2 \cdot \mathbf r_i} + 
	\bm \Delta_3 e^{i \mathbf Q_3 \cdot \mathbf r_i}.
\end{eqnarray}
Here the ordering wave vectors $\mathbf Q_1 = (2\pi, 0)$, and $\mathbf Q_{2, 3} = (-\pi, \pm \sqrt{3} \pi)$, corresponding to the mid-points of the edge of the hexagonal Brillouin zone (BZ). Moreover, the inversion-related pairs $\pm \mathbf Q_\eta$ are equivalent, as their difference is a reciprocal lattice vector.  Because the phase factors are $\exp(i \mathbf Q_\eta \cdot \mathbf r_i) = \pm 1$, it can be easily checked that the four distinct magnetic moments in the extended unit cells are $\mathbf S_{ A} = \bm\Delta_1 + \bm\Delta_2 +\bm\Delta_3$, $\mathbf S_{ B} = \bm\Delta_1 - \bm\Delta_2 -\bm\Delta_3$, $\mathbf S_{ C} = -\bm\Delta_1 + \bm\Delta_2 -\bm\Delta_3$, and $\mathbf S_{ D} = -\bm\Delta_1 - \bm\Delta_2 +\bm\Delta_3$. 
The highly symmetric tetrahedral spin order corresponds to the special case when the three vector order parameters are orthogonal to each other and have the same amplitude, i.e. $\bm \Delta_\eta = \Delta \,\hat{\mathbf e}_{\eta}$, where $\Delta$ is a real number characterizing the amplitude of spin order and $\hat{\mathbf e}_\eta$ are three mutually orthogonal unit vectors.

For electron filling fraction $n = 3/4$, this tetrahedral spin order is stabilized by a perfect Fermi surface nesting~\cite{Martin2008,Chern2012}, similar to the scenario of N\'eel order in square Kondo-lattice or Hubbard model at half filling. The Fermi surface at this filling fraction is a regular hexagon inscribed within the BZ. Pairs of parallel edges of this hexagonal Fermi surface are perfectly nested by the three ordering wave vectors $\mathbf Q_\eta$. Combined with a van Hove singularity at the corresponding Fermi energy, the magnetic susceptibility exhibits a log-square divergence $\chi(\mathbf q) \sim \log^2|\mathbf q - \mathbf Q_\eta|$ at the nesting wave vectors. As a result, the system is unstable against developing a triple-$\mathbf Q$ magnetic order which gaps out the entire Fermi surface, even for a small electron-spin coupling $J$.

The appearance of the tetrahedral spin order at $n \sim 1/4$, on the other hand, is unexpected from the Fermi surface nesting scenario. The Fermi surface in the vicinity of quarter filling has an overall circular shape, with no substantial nesting tendency. The magnetic susceptibility $\chi(\mathbf q)$, computed from the second-order perturbation expansion in $J/t$, does show a weak maximum at the three special wave vectors $\mathbf Q_\eta$ for filling fraction $n \sim 1/4$. Yet, the single, double, and triple-$\mathbf Q$ orderings remain degenerate in energy at this order. This degeneracy is lifted at the fourth order, where a positive biquadratic spin interaction $B(\mathbf S_i \cdot \mathbf S_j)^2$ with a positive coefficient $B > 0$ clearly favors the tetrahedral triple-$\mathbf Q$ order~\cite{Akagi2012}. 

Moreover, the three ordering wave vectors $\mathbf Q_\eta$ connect particular points of the Fermi surface at filling fractions $n \approx 1/4$ such that the tangents of the Fermi surface at the connected points are parallel to each other. This special geometry gives rise to a singularity analogous to the $2 k_F$ Kohn anomaly observed in an isotropic electron gas. Importantly, the positive biquadratic interaction $B$ is critically enhanced by this singularity, leading to the stabilization of the noncoplanar tetrahedral spin order~\cite{Akagi2012}. 

The noncoplanar spins in the tetrahedral order are further characterized by a discrete $Z_2$ symmetry related to the handedness of the magnetic structure. The chirality of the spin texture can be defined by the scalar product $\chi_{ijk} = \mathbf S_i \cdot \mathbf S_j \times \mathbf S_k$ of three spins on an elementary triangular loop of the lattice. A perfect tetrahedral spin order exhibits a uniform chirality $\chi_{ijk} = \pm 4 S^3 / 3 \sqrt{3}$. The two possible signs of chirality correspond to an Ising type symmetry that is spontaneously broken at low temperatures. In general, the average scalar chirality of a triple-$\mathbf Q$ order can be expressed in terms of the three magnetic order parameters as: $\chi \propto \bm\Delta_1 \cdot \bm\Delta_2 \times \bm \Delta_3$. It is clear from this expression that a nonzero chirality requires a noncoplanar configuration for the three vector order parameters. Moreover, the amplitude of the average scalar chirality is maximized in the tetrahedral spin order where the three vector order parameters are orthogonal to each other.  The two distinct chiral domains related by the $Z_2$ symmetry correspond to a right-handed and a left-handed triad of the orthogonal vectors~$\bm \Delta_\eta$.

The electronic state associated with the tetrahedral order has unusual transport and magnetoelectric properties resulting from their chiral structure. In particular, the chiral spin order exhibits a quantized Hall conductivity $\sigma_{xy} = \pm e^2/h$, with the sign determined by the scalar spin chirality~$\chi$, in the absence of applied magnetic fields~\cite{Martin2008,Chern2010}. This can be understood as follows. Because of the electron-spin coupling, the electron spins are aligned with the underlying magnetic texture. When electrons traverse a closed path in this noncoplanar spin structure, their wave function acquires a Berry phase which is half of the solid angle subtended by the electron spin as it moves around the path~\cite{ohgushi00}. For a given spin species (locally parallel or antiparallel to the magnetic texture), the effect of the Berry phase is indistinguishable from real magnetic fluxes induced by external magnetic fields that support a quantized Hall state.

It is worth noting that, while the long-range tetrahedral spin order is unstable against thermal fluctuations in 2D, the chirality order associated with a broken $Z_2$ symmetry persists at finite temperatures. Indeed, a phase transition into a chiral phase has been observed in Monte Carlo simulations of the triangular KLM at filling fractions $n\approx 1/4$~\cite{Kato2010}. Although the $Z_2$ nature of the chirality order suggests the phase transition belong to the 2D Ising universality class, Monte Carlo simulations, however, showed a clear first-order phase transition. Large-scale Langevin dynamics simulations based on KPM also observed a discontinuous onset of the triple-$\mathbf Q$ ordering at very low temperatures~\cite{Barros13}.  It is suggested that the deviation from the expected Ising universality class results from the nontrivial interplay between the discrete chiral order parameter and the underlying continuous magnetic degrees of freedom~\cite{Kato2010}. The long-range nature of electron-mediated interactions might also contribute to the unusual first-order Ising transitions.

\section{Kinetics of chiral phase ordering}

\label{sec:results}

\subsection{LLG simulations of phase ordering after a thermal quench}

\begin{figure}[t]
\centering
\includegraphics[width=1.\columnwidth]{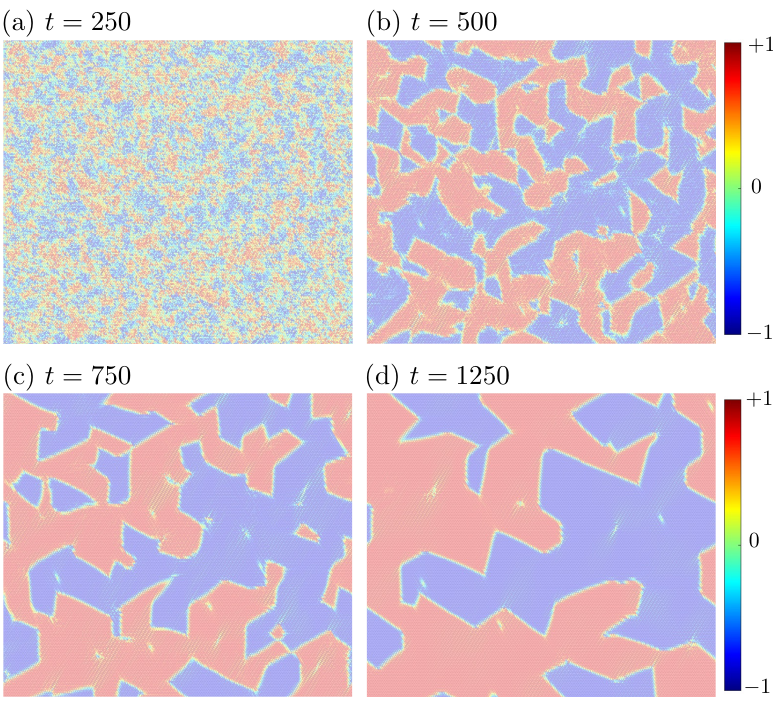}
\caption{Snapshots of local chirality parameter $\chi_{\triangle}(\mathbf r)$ at various time steps after a thermal quench of the triangular KLM at $n\sim 1/4$ filling. The color intensity indicates the local chirality normalized by its amplitude $\chi_{\rm tet} = 4 S^3 / 3 \sqrt{3}$ of the tetrahedral spin structure.  An initially random configuration is suddenly quenched to a zero temperature $T = 0$ at time $t = 0$. The ML-LLG dynamics is used to simulate the relaxation of the system toward equilibrium. }
    \label{fig:snapshots}
\end{figure}

The fact that the chiral phase transition does not belong to the expected 2D Ising universality class also suggests that the kinetics of the chiral phase transition might differ from the expected dynamical behaviors of an Ising order. For example, it is well established that the coarsening of Ising domains is characterized by a power law $L \sim t^\alpha$, where $L$ is the characteristic domain size and the growth exponent $\alpha = 1/2$ and 1/3 for a nonconserved and conserved, respectively, Ising order parameter~\cite{Bray1994,Onuki2002,Puri2009}. Since the scalar chirality is not a conserved quantity in spin dynamics, the growth of the chiral domains in KLM is expected to follow the $\alpha = 1/2$ power-law, also known as the Allen-Cahn law.

\begin{figure}
\centering
\includegraphics[width=0.99\columnwidth]{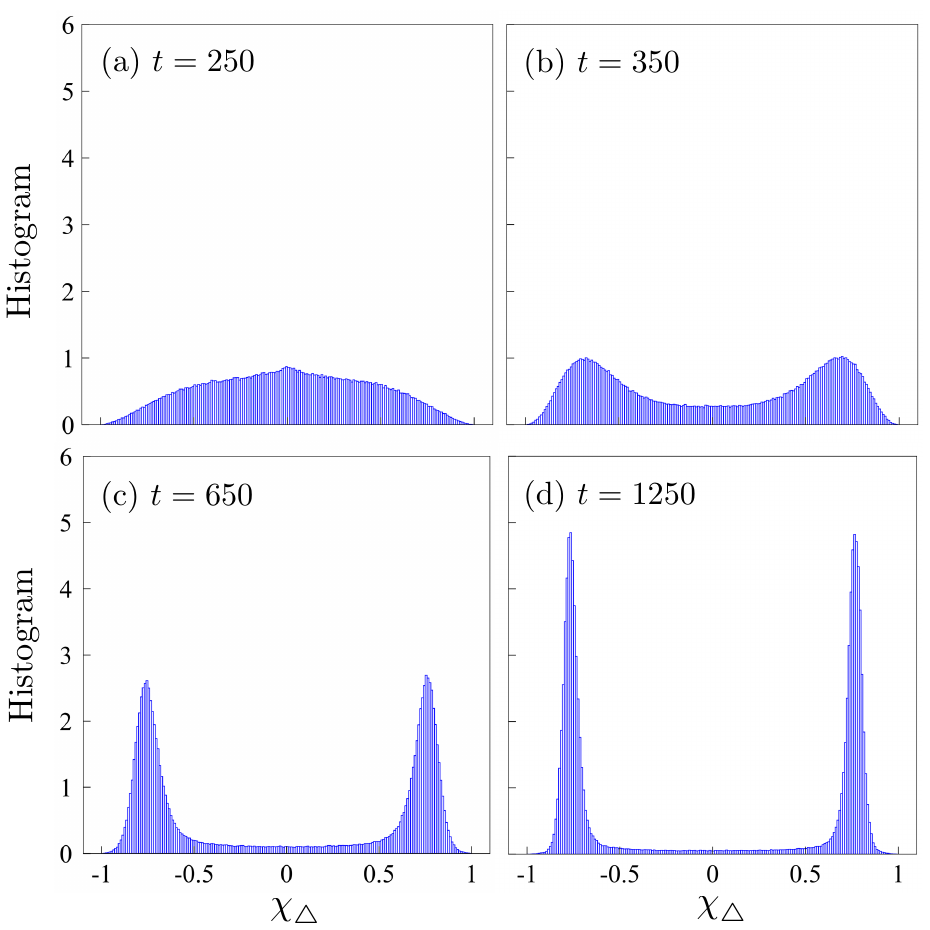}
\caption{Histogram of local spin chirality (in units of $S^3$) defined on triangular plaquettes at various times after a thermal quench. The late stage of phase ordering is characterized by a pronounced bimodal distribution with two strong peaks at $\pm \chi_{\rm tet}$, where $\chi_{\rm tet} = 4 S^3 / 3 \sqrt{3} \approx 0.77 S^3$ is the chirality of the tetrahedral spin order. }
    \label{fig:chirality-hist}
\end{figure}

To investigate the coarsening of the chiral domains, here we present further details of thermal quench simulations of the KLM at $n \sim 1/4$ filling. FIG.~\ref{fig:snapshots} shows snapshots of the local scalar chirality $\chi_{\triangle}(\mathbf r)$ at different times after a thermal quench at $t = 0$, with the color intensity indicating the value of the local chirality. At first, the random spins in the initial state give rise to a disordered distribution of local chirality as shown in FIG.~\ref{fig:snapshots}(a). The relaxation process afterward is characterized by the formation and subsequent growth of domains of a well-defined uniform chirality. At late stage, such chiral domains with a saturated $ \pm \chi_{\rm tet}$ are well established; here $ \chi_{\rm tet} = 4 S^3 / 3 \sqrt{3}$ is the scalar chirality of the tetrahedral spin order. Moreover, rather sharp interfaces, of only a few lattice constants and a vanishing chirality, separate the two domains of opposite chirality.

The formation of chiral domains is also illustrated in FIG.~\ref{fig:chirality-hist} which shows the distribution function of local spin scalar chirality $\chi_{\triangle}$ at various times after a thermal quench. At early stage of the relaxation, e.g. FIG.~\ref{fig:chirality-hist}(a), the distribution exhibits a single broad peak ranging from $-1$ to $+1$ (in units of $S^3$) with a weak maximum at $\chi = 0$. As the system further relaxes toward equilibrium, a bimodal distribution with peaks in the vicinity of $\pm \chi_{\rm tet}$ starts to emerge. At late stage of the phase ordering, the two peaks become more and more prominent as large chiral domains expand and merge with each other.

While topological defects of a broken $Z_2$ symmetry are interfaces that separate the two domains of opposite chirality, the chiral phase also exhibits $Z_2$ vortices, which are topological defects related to the underlying noncoplanar spin order~\cite{Kawamura84,Kawamura07,Rousochatzakis16}. Several examples of the $Z_2$ vortices are shown in FIG.~\ref{fig:snapshots}(c) and (d) as point-like elongated objects with vanishing chirality. Unlike the conventional integer-valued vortices in, e.g. XY spins in two dimensions, the topological number associated with a $Z_2$ vortex is either 0 or 1, which also means that topologically a vortex is equivalent to an anti-vortex. Physically, the point defect is characterized by a vortex-like structure of the vector chirality $\bm\kappa_{ijk} = (\mathbf S_i \times \mathbf S_j + \mathbf S_j \times \mathbf S_k + \mathbf S_k \times \mathbf S_i)$. The topological nature of these defects means that $Z_2$ vortices can only disappear through pair-annihilation. However, our simulations find that a $Z_2$ vortex can also be absorbed into the boundaries of chiral domains, indicating the nontrivial interplay between the two types of topological defect structures. 

\subsection{Coarsening dynamics of chiral domains}

\begin{figure}
\centering
\includegraphics[width=0.99\columnwidth]{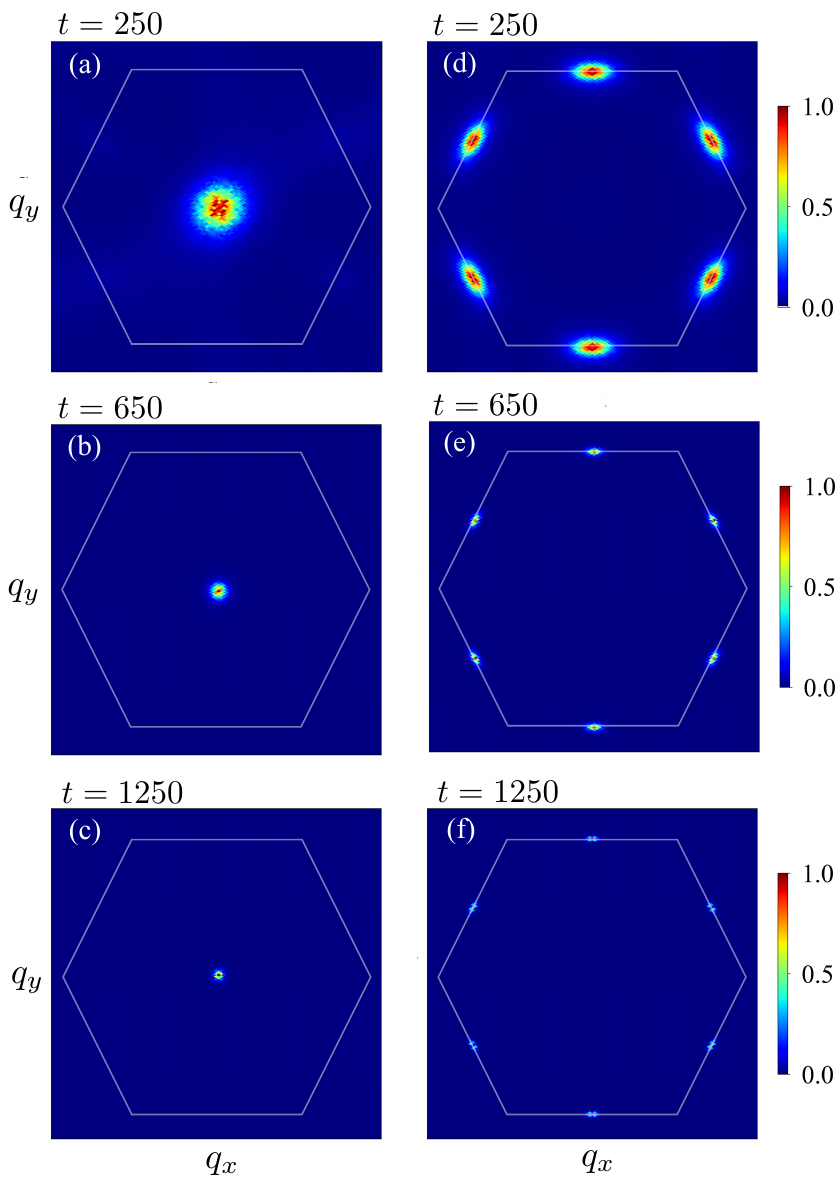}
\caption{(Left) chirality structure factor $\mathcal{S}_{\chi}(\mathbf q, t) \equiv |\tilde{\chi}(\mathbf q, t) |^2$ and (right) magnetic structure factor $\mathcal{S}_M(\mathbf q, t) \equiv |\tilde{\mathbf S}(\mathbf q, t)|^2$ at different times after a thermal quench. Compared with the circular symmetric peaks of the chirality structure factor at $\mathbf q = 0$, the six diffusive peaks of the magnetic structure factor exhibit a pronounced anisotropy. }
    \label{fig:Sk}
\end{figure}

The growth of chiral domains can also be characterized by the time-dependent structure factor of the scalar spin chirality $\mathcal{S}_{\chi}(\mathbf q, t) = |\tilde \chi(\mathbf q, t) |^2$, where $\tilde{\chi}(\mathbf q, t)$ is the Fourier transform of the local chirality variables
\begin{eqnarray}
	\tilde{\chi}(\mathbf q, t) = \frac{1}{\sqrt{N_{\triangle}}} \sum_{\mathbf r} \chi_{\triangle}(\mathbf r, t) e^{i \mathbf q \cdot \mathbf r}.
\end{eqnarray}
Here $N_{\triangle}$ is the number of triangles in the lattice and $\chi_{\triangle}(\mathbf r, t)$ denotes the scalar spin chirality defined on a local triangular plaquette centered at $\mathbf r$ and at time $t$. The chirality structure factors at different times after a thermal quench are shown in the left column of FIG.~\ref{fig:Sk}. The uniform ordering of scalar chirality gives rise to an emerging peak at wave vector $\mathbf q = 0$ in the structure factor. However, contrary to sharp Bragg peaks that are characteristic of long-range order, the quenched states here exhibit a broad diffusive peak due to the coexistence of multiple chiral domains of opposite signs, as evidenced in the snapshots shown in FIG.~\ref{fig:snapshots}. As the system equilibrates, the coarsening of chiral domains results in a brighter and sharper peak at $\mathbf q = 0$.

As discussed above, the emergence of the chiral order requires an underlying triple-$\mathbf Q$ magnetic structure. This is demonstrated by the time-dependent magnetic structure factor defined as $\mathcal{S}_M(\mathbf q, t) = |\tilde{\mathbf S}(\mathbf q, t)|^2$, where $\tilde{\mathbf S}(\mathbf q, t)$ is the Fourier transform of the spin configuration
\begin{eqnarray}
	\tilde{\mathbf S}(\mathbf q, t) = \frac{1}{\sqrt{N}} \sum_i \mathbf S_i \, e^{i \mathbf q \cdot \mathbf r_i }.
\end{eqnarray}
For a general triple-$\mathbf Q$ magnetic order given in Eq.~(\ref{eq:triple-Q}), the structure factor comprises three delta peaks: $\mathcal{S}_M(\mathbf q) = \frac{1}{3}\sum_{\eta} |\bm\Delta_\eta|^2 \,\delta(\mathbf q - \mathbf Q_\eta)$. In the special case of the symmetric tetrahedral spin order, the three delta peaks have the same magnitude. As shown in FIG.~\ref{fig:Sk}, the emerging $\mathbf q = 0$ peak in the chiral structure factor is accompanied by peaks of the magnetic structure factor at the ordering wave vectors~$\mathbf Q_{\eta}$.  Again, the multiple co-existing finite-size domains of opposite chirality after a thermal quench broaden the delta peaks into diffusive ones. Yet, contrary to the isotropic $\mathbf q = 0$ peak of the chirality structure factor, the six magnetic peaks are elongated along the edge of the BZ, indicating an anisotropy related to a directional preference of domain-walls to be discussed below. 

\begin{figure}
\centering
\includegraphics[width=0.99\columnwidth]{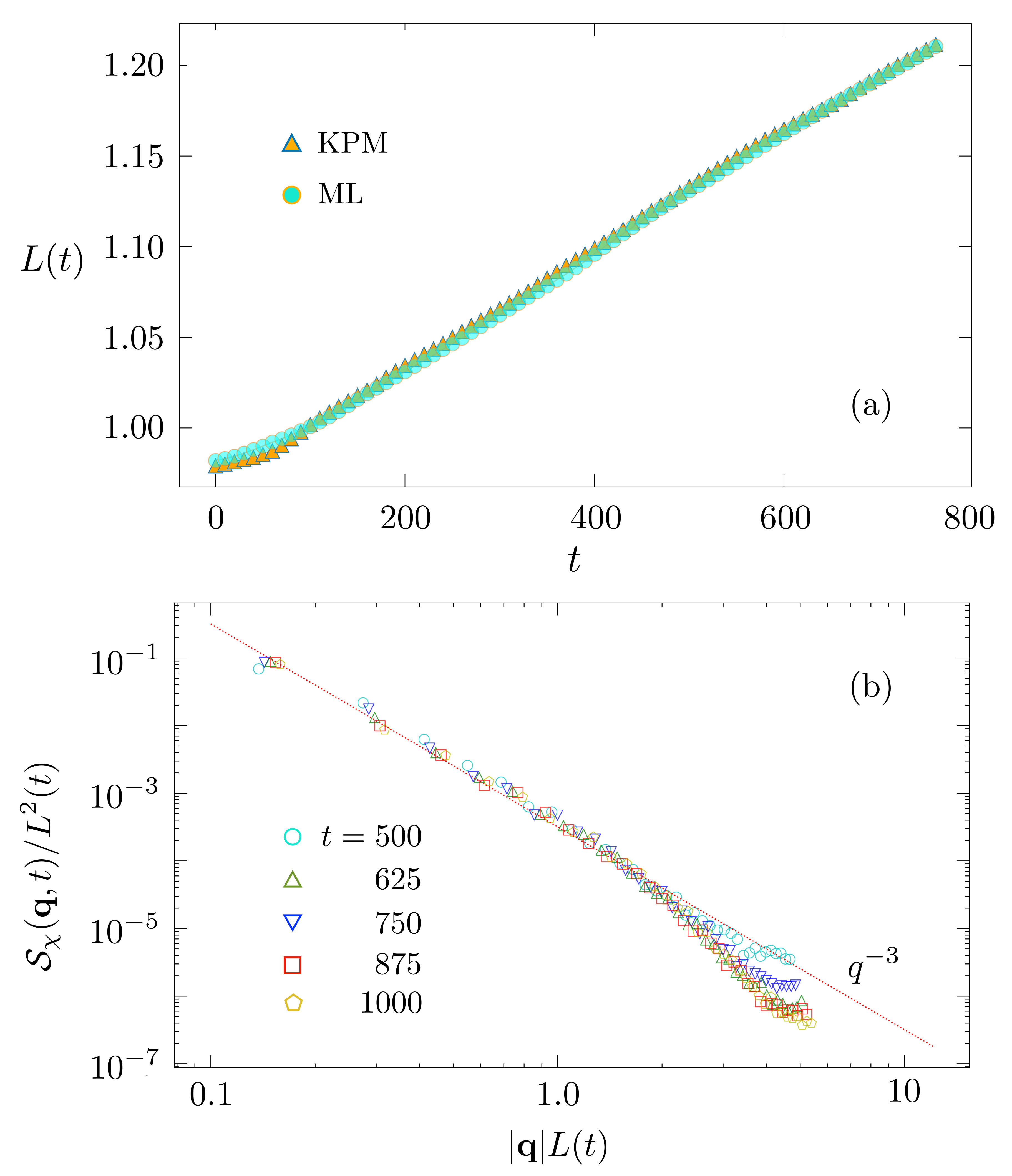}
\caption{(a) The characteristic length $L$ as a function of time extracted from the chirality structure factor. Note the linear scale for both $x$ and $y$ axes. The yellow triangles and green circles are data points obtained from LLG quench simulations using KPM and ML for the calculation of local fields. (b) Rescaled chiral structure factor $\mathcal{S}_\chi(\mathbf q, t) / L^2(t)$ versus the dimensionless wave vector $|\mathbf q| L(t)$ at different times after a thermal quench. The red dashed line indicates the $q^{-3}$ tail of the 2D Porod's law.}
    \label{fig:Lt}
\end{figure}

Formally, the width $\Delta q$ of a diffusive peak at the ordering wave vector can be used to define the correlation length of the ordered domains via the relation $L \equiv 2\pi /\Delta q$. For the chiral ordering, the time-dependent width of the peak at $\mathbf q = 0$ is computed using the chiral structure factor as a statistical weight
\begin{eqnarray}
	\Delta q(t) = 2\pi L^{-1}(t) 
	= \sum_{\mathbf q} \mathcal{S}_{\chi}(\mathbf q, t) |\mathbf q| \Big/ \sum_{\mathbf q} \mathcal{S}_{\chi}(\mathbf q, t). \quad
\end{eqnarray}
As discussed above, since the broadening of the $\mathbf q = 0$ peak in $\mathcal{S}_\chi(\mathbf q, t)$ is caused by the finite size of chiral domains, the correlation length $L$ also offers a measure of the characteristic domain size. FIG.~\ref{fig:Lt}(a) shows the time dependence of this characteristic length extracted from LLG simulations. The results obtained using ML model for the calculation of local fields are compared against those based on KPM. The agreement between the two methods again indicates that the ML force field model faithfully captures the phase ordering dynamics of the itinerant spin system.

As shown in the snapshots in FIG.~\ref{fig:snapshots}, the late-stage phase ordering is dominated by the coarsening of well-developed chiral domains, which represent ordered states of a broken $Z_2$ symmetry. Moreover, since the spin scalar chirality is not a conserved quantity of the LLG dynamics, the time evolution of the chiral Ising order is not subject to a conservation law. Phenomenologically, the coarsening dynamics of such non-conserved Ising order-parameter field $\phi(\mathbf r, t)$ is governed by the time-dependent Ginzburg-Landau (TDGL) equation~\cite{Bray1994,Onuki2002,Puri2009}: $\partial \phi / \partial t = - \Gamma \partial \mathcal{F} / \partial \phi$, where $\Gamma$ is a dissipation coefficient and $\mathcal{F}(\phi)$ is the Ginzburg-Landau free-energy functional of the order-parameter field. In particular, the growth of the Ising domains is shown to follow the Allen-Cahn law $L(t) \sim \sqrt{t}$. This square-root time dependence is also confirmed by kinetic Monte Carlo (kMC) simulations of the nearest-neighbor ferromagnetic Ising model on various lattices.

However, as shown in FIG.~\ref{fig:Lt}(a), the characteristic size $L(t)$ of chiral domains is well described by a linear time dependence for time $t > t_0$
\begin{eqnarray}
	\label{eq:linear}
	L(t) \approx L_0 + c (t - t_0), \qquad (t \gtrsim t_0).
\end{eqnarray}
Here $L_0$, $t_0$ are constants that specify the regime of linear growth, and the constant $c$ denotes the growth rate. This result is in stark contrast to the expected Allen-Cahn law for Ising-type domain coarsening. To understand this unusual result, it helps to first review the Allen-Cahn theory~\cite{Allen1972} for the dynamics of diffusive interfaces. Assuming a smooth interface, it is shown that the velocity of the interface in the normal direction is determined by its curvature: $v = \partial (\mathbf r \cdot \hat{\mathbf n})/ \partial t \big|_{\phi} = - \kappa(\mathbf r)$, where $\hat{\mathbf n}$ denotes the unit vector normal to the interface at $\mathbf r$. Approximating the interfacial velocity by the domain growth rate $v \sim dL/dt$, and the curvature by the inverse domain size $\kappa \sim 1/L$, one obtains a rate equation for the characteristic length $dL/dt \sim -1/L$. This equation can be readily integrated to give the Allen-Cahn power-law $L(t) \sim \sqrt{t}$. Coarse-grained images of Ising configurations from kMC simulations indeed confirm this curvature-driven domain growth~\cite{Bray1994,Onuki2002}. 

The above Allen-Cahn scenario, however, does not apply to the coarsening of chiral domains. As shown in FIG.~\ref{fig:snapshots}, the chiral domains are characterized by polygon-like shapes with sharp corners bounded by rather straight boundaries, instead of circular-like shapes, the characteristics of the curvature-driven coarsening.    Indeed, it has been shown that the electron-mediated effective interaction exhibits a distinct directional preference of chiral domains~\cite{Ozawa2017}. In the case of $n \sim 1/4$, the interfaces between domains of opposite chirality tend to run in directions that are perpendicular to those of nearest-neighbor bonds. The directional preferences of domain-walls also manifest themselves in the anisotropic diffusive peaks of the magnetic structure factor. As shown in FIG.~\ref{fig:Sk}(d)--(e), the six peaks at the $M$ points are elongated along the edge of the BZ boundary.   On the other hand, the domain-walls in the chiral phase of $n=3/4$ filling tend to run parallel to the directions of nearest-neighbor bonds~\cite{Ozawa2017}. Both the anisotropy itself and its dependence on the electron filling fractions underscore the important role of lattice effects as well as the complex and long-range nature of electron-mediated interactions in the coarsening dynamics of the chiral phase.  

Since the zero curvature of such straight interfaces implies a vanishing domain-wall velocity according to the Allen-Cahn equation, the growth of the chiral domains is governed by mechanisms that are beyond the TDGL $\phi^4$-theory for short-range interacting systems. The nearly linear growth Eq.~(\ref{eq:linear}) of chiral domains can be tentatively understood as follows. Even though the electron-mediated interaction is relatively longer-ranged, as discussed in Sec.~\ref{sec:BP}, one can still assume the locality of the effective interactions, which underlies the accurate approximation of the ML force-field model used here. Contrary to the Allen-Cahn scenario, information about typical domain sizes cannot be locally encoded in a straight domain-wall. As a result, when the characteristic $L$ is greater than the locality of effective interactions, the locality principle implies that the domain-wall velocity cannot be dependent on the domain structures at larger scales. Importantly, the lack of such dependences indicates a constant domain-wall velocity which is only determined by the microscopic electron model. This then implies an interfacial motion: $v \sim dL/dt = $ constant, which immediately leads to the linear time dependence in Eq.~(\ref{eq:linear}).

Finally, although the coarsening of chiral domains is dictated by the unusual linear-growth law, the chiral phase ordering nonetheless exhibits a dynamical scaling as demonstrated in FIG.~\ref{fig:Lt}(b). By properly rescaling the time-dependent structure factor and the wavevector $\mathbf q$ using the characteristic length $L(t)$, the data points at different times collapse in the vicinity of a hidden curve, indicating a scaling relation   
\begin{eqnarray}
	\label{eq:sqt}
	\mathcal{S}_{\chi}(\mathbf q, t) L^2(t) = \mathcal{G}\bigl( |\mathbf q| L(t) \bigr),
\end{eqnarray}
where $\mathcal{G}(x)$ denotes the hidden universal scaling function.  Dynamical scaling has been observed in the phase ordering of numerous Ising-type transitions. In many cases, the scaling function exhibits a $1/q^{d+1}$ power-law behavior at large wave vectors, where $d$ is the spatial dimension. This universal power-law dependence, also known as Porod's law~\cite{Bray1994, Puri2009}, can be attributed to the rather sharp interfaces that separate the two ordered states related by the $Z_2$ symmetry. For our case, the structure factor of the scalar chirality seems to be well described by the $q^{-3}$ power law for intermediate values of~$q$; see FIG.~\ref{fig:Lt}(b). The significant deviation from Porod's law at large values of the wave vector can be attributed to the complex domain-wall structures of the tetrahedral order.

\section{Conclusion and Outlook}

\label{sec:discussion}

To summarize, we have presented a scalable ML framework for the adiabatic coarsening dynamics of chiral domains in the triangular Kondo-lattice model. As the chirality order originates from the underlying tetrahedral spin order, the coarsening behaviors of chiral domains are governed by the magnetization dynamics of spins with driving forces coming from the electrons. To this end, we have generalized the Behler-Parrinello approach, a ML force-field scheme underlying most ML-based quantum molecular dynamics methods, to model the magnetization dynamics of itinerant electron systems. Assuming the locality principle for the electronic forces, a deep-learning neural network was developed to accurately encode the complex dependence of the effective magnetic field on local spin configurations. In order to preserve the symmetry of the original Hamiltonian, the group-theoretical method was employed to develop a magnetic descriptor, which is another crucial component of the BP approach.  

The ML model was trained by datasets obtained from LLG simulations where the electronic structure was solved using either exact diagonalization or kernel polynomial method. By integrating the ML force-field model with the LLG simulation, our dynamical benchmarks showed that the ML model not only accurately predicts the local magnetic fields that drive the spin dynamics, but also faithfully captures the dynamical evolution of the itinerant spin systems.

Applying our approach to study phase-ordering of the chiral phase after a thermal quench, large-scale LLG simulations enabled by the ML models uncovered an intriguing linear growth of chiral domains, which is in stark contrast to the Allen-Cahn square-root law expected for the coarsening dynamics of a non-conserved Ising order. The anomalous coarsening can be attributed to the directional anisotropy of the chiral domain-walls. The interfaces between well-developed chiral domains tend to be straight lines that are perpendicular to the directions of nearest-neighbor bonds. A combination of the locality principle and the fact that structural information at larger scales cannot be encoded in the geometry of such straight interfaces implies a constant domain-wall velocity. This result could explain the observed linear growth of chiral domains. 

The above argument, however, cannot decide the sign of the domain-wall velocity. From the phenomenological point of view, the direction of the domain-wall motion must be dictated by the geometry of the corners of the chiral domains. Consider a given vertex bounded by two interfaces, the motion of the two domain-walls are expected to shrink the area corresponding to the smaller side of the two conjugate angles. This can be viewed as the discrete version of the Allen-Cahn law. While the standard TDGL theory fails to describe the coarsening of chiral domains, an intriguing generalization is to develop a macroscopic phase-field model that takes into account the directional anisotropy of domain boundaries and the effects of vertices in determining the direction of domain-wall motions.

Also of interest is the ordering dynamics of the quasi-long-range tetrahedral spin order. The spin-spin correlations after a thermal quench are affected by three length scales: (i) the characteristic size $L$ of the chiral domains, (ii) the average distance $\ell$ between the $Z_2$ vortices, and~(iii) a characteristic length $\xi_{\rm th}$ associated with the long-wavelength thermal fluctuations \'a~la Mermin-Wanger. While the first two lengths are expected to increase with time after the quench, the thermal length $\xi_{\rm th}(T)$ is a constant which only depends on the quench temperature.  At low enough temperatures such that $\xi_{\rm th}(T) \gg L, \ell$, the spin-spin correlation length is mainly governed by the interplay of the first two length scales.  

While the characteristic size $L$ of chiral domains is found to grow linearly with time in this work, the increase of $\ell$ is related to the annihilation dynamics of $Z_2$ vortices and is shown to follow a power-law $\ell \sim t^{1/2}$~\cite{Williamson17}, similar to that of conventional vortices in XY models~\cite{Yurke93,Bray00}. In addition to pair-annihilation, as discussed above, the absorption into chiral domain-walls also contributes to the decline of $Z_2$ vortices in our case. This additional mechanism is likely to enhance the growth of $\ell$ in early stage of the phase ordering. Our simulations find that a significant number of $Z_2$ vortices remain when large chiral domains are well established. In this dynamical regime with $L \gg \ell$, the development of the tetrahedral spin order is then governed by the intrinsic pair-annihilation dynamics of $Z_2$ vortices, which will be left for future studies.

\begin{acknowledgments}
The authors thank Chen Cheng and Kotaro Shimizu for useful discussions on the implementation of descriptors and ML models. G.W.C. thanks Kipton Barros and C.~D.~Batista for collaborations on related projects. The work is supported by the US Department of Energy Basic Energy Sciences under Contract No. DE-SC0020330. The authors also acknowledge the support of Research Computing at the University of Virginia.
\end{acknowledgments}

\appendix

\section{Magnetic descriptor}

\label{sec:descriptor}

\begin{figure*}
\centering
\includegraphics[width=1.9\columnwidth]{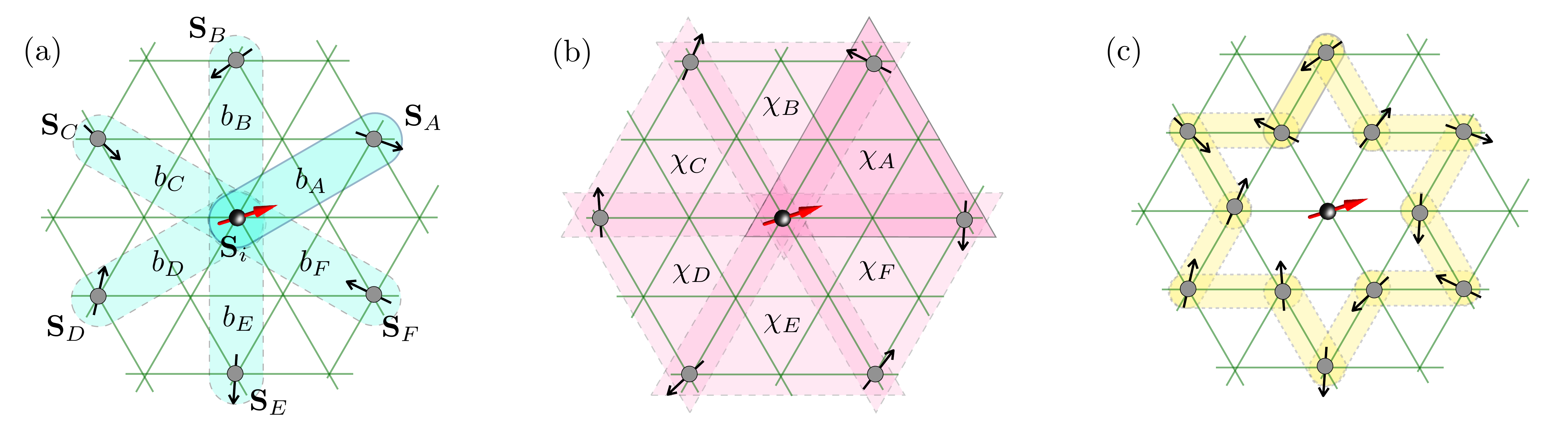}
\caption{Symmetry-invariant descriptor for the neighborhood spin configuration. (a) The six bond variables $b_k = \mathbf S_0 \cdot \mathbf S_k$, constructed from the inner product of the six third-nearest neighboring spins $\mathbf S_k$ and the center spin $\mathbf S_0$, form the basis of a four-dimensional reducible representation of the D$_6$ point group. (b) Similarly, the six scalar chirality variables $\chi_1, \chi_2, \cdots ,\chi_6$, each indicated by a triangle, also form the basis of a reducible 6-dim representation of $D_6$. The scalar chirality of a triangle $(ijk)$ is  $\chi = \mathbf S_i \cdot \mathbf S_j \times \mathbf  S_k$. (c) An example of neighboring off-site bond variables forming a 12-dim reducible representation of the $D_6$ group.}
    \label{fig:descriptor}
\end{figure*}

The magnetic descriptor developed in our previous work~\cite{zhang21} aims to provide a proper symmetry-invariant prepresentation of spin configurations in a neighborhood. As discussed in the main text, here we define a local magnetic environment centered at a lattice site-$i$ as the spin configurations within a cutoff radius $r_c$
\begin{eqnarray}
	\label{eq:Ci}
	\mathcal{C}_i = \{ \mathbf S_j \, \big| \, |\mathbf r_j - \mathbf r_i | \le r_c \}.
\end{eqnarray}
The goal here is to obtain feature variables that can faithfully distinguish different magnetic environments $\mathcal{C}_i$, yet are invariant with respect to the symmetry group of the Kondo-lattice Hamiltonian.  As the feature variables are to be used as input to the neural-network (NN) model which predicts the scalar local energy $\epsilon_i$, the symmetry requirement here is to ensure that exactly the same feature variables are obtained for any two spin configurations $\mathcal{C}_i$ and $\mathcal{C}'_i$ that are related by symmetry transformations. 

As already discussed in the main text, there are two independent symmetry groups associated with the KLM: the SU(2) or SO(3) rotation symmetry of spins and the discrete symmetry of the point group with respect to a lattice site, which in our case of the square lattice is the $D_6$ group. First, to preserve the spin-rotation symmetry, it is sufficient to introduce bond and scalar chirality variables, defined in Eq.~(\ref{eq:bond-chirality}), as building blocks for the group-theoretical computation at the second step to incorporate the lattice symmetry.  Essentially, Instead of assuming that the local energy $\epsilon_i$ depends directly on the local spin configuration $\mathcal{C}_i$, here we demand that it is a function of bond and chirality variables $\{ b_{jk}, \chi_{jkl} \}$ constructed from spins in the neighborhood. As a result, the output of the ML model is manifestly invariant under spin rotations. 

In our implementations, we set the cutoff radius $r_c = 6 a$, where $a$ is the lattice constant, for the definition of the local neighborhood in Eq.~(\ref{eq:Ci}). Let $M$ be the number of spins included in the neighborhood $\mathcal{C}_i$, the number of bond and scalar chirality variables scales roughly as $\mathcal{O}(M^2)$ and $\mathcal{O}(M^3)$, respectively. Due to the large number of such scalar variables within the neighborhood, we restrict the distance between pairs of neighboring spins to within another cutoff distance~$r_c' = 2a$ for practical implementations.

Next, we consider the inclusion of the lattice symmetry.  A general approach to construct invariants of discrete symmetry groups is based on the so-called bispectrum method~\cite{kondor07,bartok13}, which are the triple product of the basis functions (or coefficients) of the irreducible representations (IRs) of the symmetry group~\cite{hamermesh62}. To obtain all the IR coefficients of the magnetic environment, we consider the finite {\em reducible} representation given by the bond and scalar chirality variables $\{ b_{jk}, \chi_{jkl} \}$ discussed above.
While there are standard procedures to decompose a high-dimensional reducible representation, our task here is considerably simplified by noting that the representation matrices in terms of the bond/chirality basis are automatically block-diagonalized. This is because the distance between a spin-pair (bond) or spin-triplet (scalar chirality) from the center site is preserved by operations of the discrete point group. Each block then corresponds to $b_{jk}$ or $\chi_{jkl}$ of a fixed distance; see FIG.~\ref{fig:descriptor}. 

In the case of a triangular lattice, the dimension of each block is either 6 or 12. Take for example the six third-nearest neighboring spins $\mathbf S_A, \mathbf S_B, \cdots, \mathbf S_F$ in FIG.~\ref{fig:descriptor}(a). The resultant six bond variables $b_\mu = \mathbf S_i \cdot \mathbf S_\mu$ ($\mu = A, B, \cdots, F)$ then form the basis of a dimension-6 representation of the D$_6$ group, and can be decomposed into $6 = A_1 \oplus B_2 \oplus E_1 \oplus E_2$, with the following basis~\cite{hamermesh62}:
\begin{equation*} \label{eq1}
\begin{split}
f^{A_1} & = b_{A} + b_{B} + b_{C} + b_{D} + b_E + b_F \\
f^{B_2} & = b_{A} - b_{B} + b_{C} - b_{D} + b_E - b_F \\
f^{E_1}_x & = \frac{\sqrt{3}}{2} ( b_{B} + b_{C} - b_E - b_F) \\
f^{E_1}_y & = \frac{1}{2} ( 2b_A + b_B - b_C - 2 b_D  -b_E + b_F) \\
f^{E_2}_x & = \frac{1}{2} ( 2b_A - b_B - b_C + 2 b_D - b_E - b_F) \\
f^{E_2}_y & = \frac{\sqrt{3}}{2} (-b_B + b_C - b_E + b_F)
\end{split}
\end{equation*} 
Another example is the 6-dimensional reducible representation based on the six scalar chirality variables $\chi_A, \chi_B$, $\cdots$, $\chi_F$ that depend on the center spin $\mathbf S_i$ and 6 neighboring spins as shown in FIG.~\ref{fig:descriptor}(b). In this case, since the scalar chirality variables change sign under an odd permutation of the three spins, their decomposition is given by $6 = A_2 \oplus B_1 \oplus E_1 \oplus E_2$ with the basis 
\begin{equation*} \label{eq2}
\begin{split}
f^{A_2} & = \chi_{A} + \chi_{B} + \chi_{C} + \chi_{D} + \chi_E + \chi_F \\
f^{B_1} & = \chi_{A} - \chi_{B} + \chi_{C} - \chi_{D} + \chi_E - \chi_F \\
f^{E_1}_x & = \frac{\sqrt{3}}{2} ( \chi_{B} + \chi_{C} - \chi_E - \chi_F) \\
f^{E_1}_y & = \frac{1}{2} ( 2 \chi_A + \chi_B - \chi_C - 2 \chi_D  - \chi_E + \chi_F) \\
f^{E_2}_x & = \frac{1}{2} ( 2 \chi_A - \chi_B - \chi_C + 2 \chi_D - \chi_E - \chi_F) \\
f^{E_2}_y & = \frac{\sqrt{3}}{2} (- \chi_B + \chi_C - \chi_E + \chi_F)
\end{split}
\end{equation*}
Finally, we give an example of a subspace of 12 dimensions, shown in FIG.~\ref{fig:descriptor}(c). In this case, the basis of this reducible representation is given by the off-site bond variables (which means the center spin is not involved) from 12 neighboring spins. The decomposition of this 12-dimensional representation is $12 = A_1 \oplus A_2 \oplus B_1 \oplus B_2 \oplus 2\, E_1 \oplus 2\, E_2$.

By repeating the same procedures for all invariant blocks, one obtains all the IRs of the bond/chirality variables in the neighborhood $\mathcal{C}_i$.  For convenience, we arrange the basis functions of a given IR in the decomposition into a vector $\bm f^{(\Gamma, r)} = \left(f^{(\Gamma, r)}_1, f^{(\Gamma, r)}_2, \cdots, f^{(\Gamma, r)}_{n_\Gamma} \right)$ where $\Gamma$ labels the IR, $r$ enumerates the multiple occurrences of IR $\Gamma$ in the above decomposition, and $n_{\Gamma}$ is the dimension of the IR. Given these basis functions, one can immediately obtain a set of invariants called power spectrum $\{p^\Gamma_r\}$ defined in Eq.~(\ref{eq:power-spectrum}). However, as discussed in the main text, feature variables based only on the power spectrum are incomplete in the sense that the relative phases between different IRs are ignored. For example, the relative ``angle" between two IRs of the same type: $\cos\theta = (\bm{f}^{\Gamma}_{r_1}\cdot\bm{f}^{\Gamma}_{r_2})/|\bm{f}^{\Gamma}_{r_1}||\bm{f}^{\Gamma}_{r_2}|$ is also an invariant of the symmetry group. Without such phase information, the NN model might suffer from additional error due to the spurious symmetry, namely two IRs can freely rotate independently of each other.

A systematic approach to include all relevant invariants, including both amplitudes and relative phases, is the bispectrum method~\cite{kondor07,bartok13}. Given the coefficients $\bm{f}^\Gamma_r$ of all IRs, the bispectrum coefficients are given by
\begin{eqnarray}
	B^{\Gamma, \Gamma_1, \Gamma_2}_{r, r_1, r_2} = \sum_{\alpha\beta\gamma} C^{\Gamma; \Gamma_1, \Gamma_2}_{\alpha, \beta,\gamma} f^{\Gamma}_{r, \alpha} f^{\Gamma_1}_{r_1, \beta} f^{\Gamma_2}_{r_2, \gamma}
\end{eqnarray}
where $C^{\Gamma; \Gamma_1, \Gamma_2}$ are the Clebsch-Gordan coefficients of the point group~\cite{kondor07}. The bispectrum coefficients are triple products of coefficients of the same or different IRs, similar to the scalar product $\mathbf a \cdot \mathbf b \times \mathbf c$ of three vectors.   When one of the three IRs, say $\Gamma$, is the trivial identity IR, the resultant bispectrum coefficient is reduced to the inner product of two IRs of the same type, i.e. they transform in exactly the same way under the symmetry operations. The resultant invariants are similar to the inner product $\mathbf a \cdot \mathbf b$ of two vectors. 

However, the number of bispectrum coefficients is often too large for practical applications, and some of them are redundant. Here we have implemented a descriptor that is modified from the bispectrum method~\cite{zhang21}. We introduce the reference basis functions $\bm f^{\Gamma}_{\rm ref}$ for each IR of the point group. These reference bases are computed by averaging large blocks of bond and chirality variables, such that they are less sensitive to small changes in the neighborhood spin configurations. We then define the relative ``phase" of an IR as the projection of its basis functions onto the reference basis defined in Eq.~(\ref{eq:IR-phase}).

The relative phase between two IRs $\Gamma_1$ and $\Gamma_2$ can then be restored from their relative phase with the corresponding reference IR coefficients $\bm f^{\Gamma_1}_{\rm ref}$ and $\bm f^{\Gamma_2}_{\rm ref}$, and a few additional bispectrum coefficients of these reference IRs. 
The effective coordinates are then the collection of power spectrum coefficients and the relative phases: $\{ G_\ell \} = \{ p^{\Gamma}_r \,\, , \,\, \eta^\Gamma_r \}$.  These feature variables, which are invariant with respect to both the SO(3) spin rotation symmetry and the point group site-symmetry of the lattice, are input to the neural network; see FIG.~\ref{fig:ml-scheme}.  The local energy function $\epsilon_i$ given by the output of the NN depends only on these effective coordinates, hence obviously preserving the symmetry of the original KLM.

\section{Neural network model}

\label{sec:nn-model}

The neural network is implemented using PyTorch~\cite{Paszke2019}. The NN model comprises eight hidden layers with $2048\times1024\times512\times256\times128\times64\times64\times64$ neurons respectively. The input layer of the model was determined by the number of feature variables $\{G^\Gamma_r\}$, which in this case is 1806. There is only one neural at the output, which is utilized to predict the local energy. Rectified linear unit (ReLU)~\cite{Nair2010} activation function is employed between layers of the NN model. Given that the torque $\textbf{T}_i=\textbf{S}_i\times\textbf{H}_i$ is the key component for deriving the Landau-Lifshitz-Gilbert (LLG) dynamics, the mean square error (MSE) of both the total energy and the local effective fields are used as the loss function
\begin{equation}
	\mathcal{L}=\eta_H\sum^N_{i=1} \bigl| \mathbf T_i - \hat{\mathbf T}_i \bigr|^2
	+ \eta_E \Bigl| E - \sum_i^N \hat{\epsilon}_i \Bigr|^2. \,\,
	\label{eq:loss_func}
\end{equation}
Here quantities with a hat denote predictions from the ML models. The parameters $\eta_H$ and $\eta_E$ are introduced to control the relative weight of force and energy predictions for training the ML models. In our case, we use a relatively large $\eta_H$ to emphasize the accuracy of force predictions. 
The optimization of the NN parameters is carried out utilizing the Adam stochastic optimization algorithm~\cite{Kingma2017} with a learning rate that undergoes exponential decay as the number of training iterations increases, starting at 0.1 and decreasing to 0.0001. A total of 15,000 snapshots of spins and the corresponding local fields, obtained from 30 independent KPM-LLG simulations, are used as the training (90\%) and testing (10\%) datasets. The training is conducted over 200 epochs. In order to mitigate overfitting, a 5-fold cross-validation strategy is employed~\cite{Stone1974}.

\bibliography{ref}

\end{document}